\documentstyle[preprint,eqsecnum,aps]{revtex}
\tightenlines                                                                                                                                                                
\newcommand{\be}{\begin{equation}}
\newcommand{\ee}{\end{equation}}
\newcommand{\bea}{\begin{eqnarray}}
\newcommand{\eea}{\end{eqnarray}}
\newcommand{\ba}{\begin{array}}
\newcommand{\ea}{\end{array}}
\newcommand{\ep}{\epsilon}
\newcommand{\Th}{\Theta}

\newcommand{\la}{\lambda}

\newcommand{\de}{\delta}

\newcommand{\pa}{\partial}
\newcommand{\pax}{\partial_x}

\newcommand{\no}{\nonumber}

\newcommand{\res}{\mbox{res}}
\newcommand{\tr}{\mbox{tr}}
\newcommand{\Tr}{\mbox{Tr}}

\newcommand{\lb}{\label}
\newcommand{\Q}{\hat{Q}}
\newcommand{\tQ}{\tilde{Q}}
\newcommand{\vphi}{\varphi}
\newcommand{\al}{\alpha}
\newcommand{\bt}{\beta}

\begin{document}

\title{Matrix Formulation of Hamiltonian Structures of \\
Constrained KP Hierarchy}

\author{Wen-Jui Huang$^1$, Jiin-Chang Shaw$^2$ and Ming-Hsien Tu$^3$ }
\address{
$^1$ Department of Physics, National Changhua University of Education, \\
Changhua, Taiwan\\
$^2$ Department of Applied Mathematics, National Chiao Tung University, \\
Hsinchu, Taiwan, \\
and\\
$^3$ Department of Physics, National Tsing Hua University, \\
Hsinchu, Taiwan
}
\date{\today}
\maketitle

\begin{abstract}
We give a matrix formulation of the Hamiltonian structures of constrained KP
hierarchy. First, we derive from the matrix formulation the Hamiltonian structure
of the one-constraint KP hierarchy, which was originally obtained by 
Oevel and Strampp. We then generalize the derivation to the multi-constraint case
and show that the resulting bracket is actually the second Gelfand-Dickey
bracket associated with the corresponding Lax operator. The matrix formulation
of the Hamiltonian structure of the one-constraint KP hierarchy in the form
introduced in the study of matrix model is also discussed
\end{abstract}

\newpage

\section{Introduction}

The Gelfand-Dickey (GD) hierarchy is defined by\cite{D1}
\be
l_n=\pa^n+u_{n-1}\pa^{n-1}+\cdots+u_0
\lb{laxln}
\ee
which satisfies the hierarchy equations
\be
\frac{\pa l_n}{\pa t_k}=[(l_n)_+^{k/n},l_n]
\lb{eqln}
\ee
Here $A_{\pm}$ denote the differential part and the integral part of the 
pseudo-differential operator $A$. The second Hamiltonian structure
of (\ref{laxln}) is described by the GD bracket which in operator
form can be written as
\bea
\Th^{GD}_2(\frac{\de H}{\de l_n})&\equiv&\{l_n,H\}
\equiv\{u_{n-1},H\}\pa^{n-1}+
\{u_{n-2},H\}\pa^{n-2}+\cdots+\{u_0,H\}\no\\
&=&(l_n\frac{\de H}{\de l_n})_+l_n-l_n(\frac{\de H}{\de l_n}l_n)_+
\lb{gd2}
\eea
where
\be
\frac{\de H}{\de l_n}\equiv\pa^{-1}\frac{\de H}{\de u_0}+
\cdots+\pa^{-n}\frac{\de H}{\de u_{n-1}}
\ee
Note that, from (\ref{eqln}), the equation for $u_1$ is trivial thus we can set $u_1=0$.
However, imposing such a constraint leads to a modification of the GD bracket 
(\ref{gd2}) due to
the Dirac reduction. The modified bracket is found to be
\be
\bar{\Th}^{GD}_2(\frac{\de H}{\de l_n})=
(l_n\frac{\de H}{\de l_n})_+l_n-l_n(\frac{\de H}{\de l_n}l_n)_++
\frac{1}{n}[l_n,\int\res[l_n,\frac{\de H}{\de l_n}]]
\lb{mgd2}
\ee
where $u_1$ in $l_n$ and in $\de H/\de l_n$ are
both set to zero and $\res(\sum_ia_i\pa^i)\equiv a_{-1}$.

The above description of the GD hierarchy is relied on the use of fractional-power
pseudo-differential operators \cite{GD} associated with the scalar Lax operator $l_n$. 
However, one can put the formalism on a more general setting by considering 
the GD hierarchy as a reduction of a system of $n$ first-order equations. 
Following Dickey \cite{D2}, one can substitute for the scalar Lax operator $l_n$
a $n\times n$ first order differential operator
\bea
{\cal{L}}&=&I\pa+U\qquad \pa\equiv \pa/\pa x\\
U&=&
\left(
\ba{ccccc}
 0 & -1 & 0 &\cdots  & \cdots \\
\cdots  & \cdots & -1 & \cdots & \cdots\\
  \cdots &\cdots &\cdots &\cdots & 0\\
 \cdots&\cdots &\cdots & 0 & -1\\
 u_0 & u_1 & \cdots & \cdots &u_{n-1}
\ea
\right)
\eea
This is the basis of the extension to the more general situation when
the matrix $U$ belongs to a semi-simple Lie algebra, 
a task that was accomplished by Drinfeld and Sokolov \cite{DS}.
Using matrix notation, the hierarchy equations (\ref{eqln}) can be expressed as
\be
\pa_t{\cal{L}}=[Q,{\cal{L}}]
\lb{eqmatrix}
\ee
where $Q$ is another $n\times n$ matrix which can not be arbitrarily chosen.
We have to choose properly the matrix $Q$ such that $[Q,{\cal{L}}]$ consistent
with the form of $\pa_t{\cal{L}}$. In fact, in view of (\ref{eqmatrix}), if the last column
of $Q$ is given then the rest of the elements of $Q$ can be fixed.
Following this strategy, it has been shown \cite{D2} that the Hamiltonian structure 
of the GD hierarchy  can be extracted 
from (\ref{eqmatrix}) and turns out to be the GD bracket (\ref{gd2}).

Recently, there has much interest in the so-called constrained 
KP hierarchy \cite{C1,XL,SS,OS,C2,Y,BX1,D3,AGZ} 
which has a pseudo-differential Lax operator of the form
\be
L_{(n,m)}=\pa^n+u_{n-1}\pa^{n-1}+\cdots+u_0+
\sum_{k=1}^m\phi_k\pa^{-1}\psi_k
\ee
and satisfies the evolution equation
\be
\frac{\pa L_{(n,m)}}{\pa t_k}=[(L_{(n,m)})_+^{k/n},L_{(n,m)}]
\lb{eqckp}
\ee
From (\ref{eqckp}), it can be shown that each $\phi_i$ (each $\psi_i$) is an eigenfunction
(adjoint eigenfunction) of the constrained KP hierarchy, i.e.
\be
\frac{\pa \phi_i}{\pa t_k}=((L_{(n,m)})_+^{k/n}\phi_i)\qquad
\frac{\pa \psi_i}{\pa t_k}=-((L_{(n,m)}^*)_+^{k/n}\psi_i)
\ee
The bi-Hamiltonian structure of (\ref{eqckp}) have been 
constructed by Oevel and Strampp \cite{OS}
(see also \cite{C2}).  The matrix formulation of the cKP hierarchy
and its Hamiltonian structure have been discussed in refs.\cite{AGZ,AFGZ}.
In contrast to the affine Lie algebraic approach \cite{AGZ,AFGZ}, 
in this paper we shall follow Dickey's approach \cite{D2} to give 
an elementary derivation of the Hamiltonian structure associated with the Lax operator
$L_{(n,m)}$. In our approach, the Hamiltonian structures obtained by Oevel and
Strampp \cite{OS} come out quite naturally.
Since the first Hamiltonian structure can be obtained from the second Hamiltonian
structure by replacing the Lax operator $L_{(n,m)}$ by $L_{(n,m)}+\la$,
where $\la$ is called the spectral parameter, we shall focus only on the second
structure.

Our paper is organized as follows: In Sec. II, we follow the approach close to that
of \cite{D2} to build up the matrix equation (\ref{eqmatrix}) associated 
with the Lax operator $L_{(n,1)}$ and then derive the Hamiltonian structure which
was obtained by Oevel and Strampp. In Sec. III, we generalize this formulation to
the Lax operator $L_{(n,m)}$ and obtain its Hamiltonian structure. We further
show that this Hamiltonian structure is, in fact, the GD bracket (\ref{gd2}) defined by the
Lax operator $L_{(n,m)}$. In Sec. IV, we work out a few simple examples explicitly.
Concluding remarks are presented in the Sec. V.

\section{Matrix formulation of Oevel and Strampp's Hamiltonian structure}
Now we consider the matrix formulation corresponding to the
Lax operator for the one-constraint KP hierarchy
\be
L_{(n,1)}=\pa^n+u_{n-1}\pa^{n-1}+\cdots+u_0+\phi\pa^{-1}\psi.
\lb{laxln1}
\ee
A matrix representation can be easily found by expressing the constraint equation
$L_{(n,1)}\vphi=0$ in a matrix form as ${\cal{L}}_{(n,1)}\Phi=0$, 
where ${\cal{L}}_{(n,1)}$ is a square matrix and $\Phi$ is a column matrix.
An ansatz is given by
\bea
{\cal{L}}_{(n,1)}&=&I\pa+U\qquad \pa\equiv \pa/\pa x
\lb{matrixln1}\\
U&=&
\left(
\ba{cccccc}
0 & -\psi & 0 & 0 & \cdots & 0 \\
\cdots & 0 & -1 & 0 &\cdots  & \cdots \\
\cdots &\cdots  & \cdots & -1 & \cdots & \cdots\\
 \cdots& \cdots &\cdots &\cdots &\cdots & 0\\
\cdots& \cdots&\cdots &\cdots & 0 & -1\\
\phi & u_0 & u_1 & \cdots & \cdots &u_{n-1}
\ea
\right)
\eea
where $U$ is a $(n+1)\times(n+1)$ matrix and the numeration 
of rows and columns will be $i, j=-1,0,1,2,\cdots,n-1$.

We introduce another $(n+1)\times(n+1)$ matrix $Q$ and ask the commutator
$[Q,{\cal{L}}_{(n,1)}]\equiv N$ to be consistent with the form of $\pa_t{\cal{L}}_{(n,1)}$. 
Let's now compute the matrix elements of the commutator:

(1) $i=-1$, or  $j=-1$
\bea
N_{-1,-1}&=&-Q'_{-1,-1}+\phi Q_{-1,n-1}+\psi Q_{0,-1}
\lb{n1}\\
N_{-1,i}&=&-Q'_{-1,i}-\ep_iQ_{-1,i-1}+\psi Q_{0,i}+u_iQ_{-1,n-1}\qquad
(\ep_0=\psi,\ep_1=\cdots=\ep_{n-2}=1)
\lb{n2}\\
N_{i,-1}&=&-Q'_{i,-1}+\phi Q_{i,n-1}+Q_{i+1,-1}\qquad (i=0,1,2,\cdots,n-2)
\lb{n3}\\
N_{n-1,-1}&=&-Q'_{n-1,-1}+\phi Q_{n-1,n-1}-\phi Q_{-1,-1}-\sum_{j=0}^{n-1}
u_jQ_{j,-1}
\lb{n4}
\eea

(2) $i\neq -1$ and $j\neq -1$
\bea
N_{i,0}&=&-Q'_{i,0}-\psi Q_{i,-1}+u_0Q_{i,n-1}+Q_{i+1,0}
\lb{n5}\\
N_{n-1,0}&=&-Q'_{n-1,0}-\psi Q_{n-1,-1}+u_0Q_{n-1,n-1}-\phi Q_{-1,0}
-\sum_{j=0}^{n-1}u_jQ_{j,0}
\lb{n6}\\
N_{i,j}&=&-Q'_{i,j}-Q_{i,j-1}+u_jQ_{i,n-1}+Q_{i+1,j}
\lb{n7}\\
N_{n-1,j}&=&-Q'_{n-1,j}-Q_{n-1,j-1}+u_jQ_{n-1,n-1}-\phi Q_{-1,j}-
\sum_{k=0}^{n-1}u_kQ_{k,j}
\lb{n8}
\eea
where $i=0,1,\cdots,n-2$ and $j=1,2,\cdots,n-2$.

In order to make $N$ of the form $\pa_t{\cal{L}}_{(n,1)}$, we write
$N=\de {\cal{L}}_{(n,1)}$; i.e.
\bea
N_{-1,0}&=& -\de \psi 
\lb{dynapsi}\\
N_{n-1,-1}&=&\de \phi  
\lb{dynaphi}\\
N_{n-1,j}&=& \de u_j \qquad (j=0,1,\cdots,n-1)
\lb{dynau}
\eea
and $N_{i,j}=0$ for other $(i,j)$. In the matrix formulation the equation
$\de {\cal{L}}_{(n,1)}=[Q, {\cal{L}}_{(n,1)}]$ would serve as a definition
of the Hamiltonian structure. In other words, 
$\de \psi$, $\de \phi$ and $\de u_j$ are to be identified as
$\{\psi, H\}$, $\{\phi, H\}$ and $\{u_j, H\}$, respectively. 
Of course, certain matrix elements of $Q$ must be
identified as the components of the gradient of $H$. 
However we shall do such an identification later.

We shall see that one can solve $Q_{i,j}$ in terms of
$Q_{0,-1}$ and $Q_{i,n-1}$ ($i=-1,0,\cdots,n-1$). To this end we introduce
\bea
\hat{Q}_i &\equiv&\sum_{k=0}^{n-1}Q_{ik}\pa^k\qquad 
\hat{Q}^j\equiv\sum_{k=0}^{n-1}\pa^{-1-k}Q_{kj}
\lb{defq}\\
\hat{N}_i&\equiv& \sum_{k=0}^{n-1}N_{ik}\pa^k 
\lb{defn}
\eea
then (\ref{n5}) and (\ref{n7}) is equivalent to
\be
\hat{N}_i=-\pa\hat{Q}_i-\psi Q_{i,-1}+\hat{Q}_{i+1}+
Q_{i,n-1}(L_{(n,1)})_+\qquad (i=0,1,\cdots,n-2)
\lb{n}
\ee
While (\ref{n6}) and (\ref{n8}) give
\be
\hat{N}_{n-1}=-\pa\hat{Q}_{n-1}-\psi Q_{n-1,-1}-\phi\hat{Q}_{-1}+
Q_{n-1,n-1}(L_{(n,1)})_+-\sum_{k=0}^{n-1}u_k\hat{Q}_k
\ee
Since $\hat{N}_i=0$  $(i=0,1,\cdots,n-2)$ we deduce a recursion relation for $\hat{Q}_i$
from (\ref{n})
\be
\hat{Q}_{i+1}=\pa\hat{Q}_i-Q_{i,n-1}(L_{(n,1)})_++\psi Q_{i,-1}\qquad (i=0,1,\cdots,n-2)
\lb{recursive}
\ee
Define
\be
\hat{Q}_n\equiv -\sum_{k=0}^{n-1}u_k\hat{Q}_k-\hat{N}_{n-1}-\phi\hat{Q}_{-1}
\lb{qn}
\ee
Then (\ref{recursive}) also holds for $i=n-1$. Hence
\bea
\Q_1&=&\pa\Q_0-Q_{0,n-1}(L_{(n,1)})_++\psi Q_{0,-1}\\
\Q_2&=&\pa\Q_1-Q_{1,n-1}(L_{(n,1)})_++\psi Q_{1,-1}\no\\
&=&\pa^2\Q_0-\pa Q_{0,n-1}(L_{(n,1)})_+-Q_{1,n-1}(L_{(n,1)})_++
\pa\psi Q_{0,-1}+\psi Q_{1,-1}\\
&\cdots&\no\\
\Q_i&=&\pa^i(\Q_0-\sum_{k=0}^{i-1}\pa^{-1-k}Q_{k,n-1}(L_{(n,1)})_++\sum_{k=0}^{i-1}
\pa^{-1-k}Q_{k,-1}\psi)\\
&\cdots&\no\\
\Q_n&=&\pa^n(\Q_0-\sum_{k=0}^{n-1}\pa^{-1-k}Q_{k,n-1}(L_{(n,1)})_++
\sum_{k=0}^{n-1}\pa^{-1-k}Q_{k,-1}\psi)
\eea
By the virtue of (\ref{qn}), $\Q_n$ must be a differential operator of order $\leq n-1$ it follows
\bea
& &(\Q_0-\sum_{k=0}^{n-1}\pa^{-1-k}Q_{k,n-1}(L_{(n,1)})_++\sum_{k=0}^{n-1}
\pa^{-1-k}Q_{k,-1}\psi)_+\no\\
& &=(\Q_0-\sum_{k=0}^{n-1}\pa^{-1-k}Q_{k,n-1}(L_{(n,1)})_+)_+=0
\eea
Define
\be
X\equiv\sum_{k=0}^{n-1}\pa^{-1-k}Q_{k,n-1}=\Q^{n-1}
\ee
we get
\be
\Q_0=(X(L_{(n,1)})_+)_+=(XL_{(n,1)})_+
\lb{q0}
\ee
and 
\be
\Q_i=\pa^i(XL_{(n,1)})_+-(\pa^iX)_+(L_{(n,1)})_++(\pa^i\Q^{-1})_+\psi
\ee
From (\ref{qn})
\bea
\hat{N}_{n-1}&=&-\pa^n(XL_{(n,1)})_++(\pa^nX)_+(L_{(n,1)})_+-
(\pa^n\Q^{-1})_+\psi-\phi\Q_{-1}\no\\
& &-\sum_{k=0}^{n-1}u_k[\pa^k(XL_{(n,1)})_+-(\pa^kX)_++(\pa^k\Q^{-1})_+\psi]\no\\
&=&-(L_{(n,1)})_+(XL_{(n,1)})_++(L_{(n,1)}X)_+(L_{(n,1)})_+-
(L_{(n,1)}\Q^{-1})_+\psi-\phi\Q_{-1}
\eea
In view of $\de u_j=N_{n-1,j}$ we have
\be
\de (L_{(n,1)})_+=(L_{(n,1)}X)_+(L_{(n,1)})_+-(L_{(n,1)})_+(XL_{(n,1)})_+-
(L_{(n,1)}\Q^{-1})_+\psi-\phi\Q_{-1}
\lb{lplus}
\ee
It remains to characterize $\Q_{-1}$ and $\Q^{-1}$. 
Since $N_{i,-1}=0$ $(i=-1,0,\cdots,n-2)$, (\ref{n1}) and (\ref{n3}) give
\bea
Q_{-1,-1}&=&\int^x(\phi Q_{-1,n-1}+\psi Q_{0,-1})
\lb{q-1}\\
Q_{i+1,-1}&=&Q'_{i,-1}-\phi Q_{i,n-1}\qquad (i=0,1,\cdots,n-2)
\lb{qi+1}
\eea
If we define
\be
Q_{n,-1}\equiv -\phi Q_{-1,-1}-\sum_{k=0}^{n-1}u_kQ_{k,-1}-N_{n-1,-1}
\lb{qn-1}
\ee 
(\ref{n4}) becomes (\ref{qi+1}) with $i=n-1$. It follows that
\be
(\pa^{-n}Q_{n,-1}+\cdots+\pa^{-1}Q_{1,-1})=(\Q^{-1})'-\Q^{n-1}\phi.
\ee
Using $(\Q^{-1})'=[\pa,\Q^{-1}]$ we get
\be
Q_{n,-1}=\pa^n(Q_{0,-1}-\Q^{-1}\pa-X\phi)
\ee
Since $Q_{n,-1}$ is a scalar, it follows
\be
[\pa^n(Q_{0,-1}-\Q^{-1}\pa-X\phi)\pa^{-1}]_+=0
\ee
Moreover, since $\pa^n\Q^{-1}$ is a pure differential operator we have
\be
\pa^n\Q^{-1}=[\pa^n(Q_{0,-1}-X\phi)\pa^{-1}]_+
\lb{diff}
\ee
Next from (\ref{n2})
\be
Q_{-1,i-1}=-Q'_{-1,i}+\psi Q_{0,i}+u_iQ_{-1,n-1}\qquad (i=1,2,\cdots,n-1)
\lb{key}
\ee
Define $\tQ_{-1,-1}=-Q'_{-1,0}+\psi Q_{0,0}+u_0Q_{-1,n-1}$ then
in view of (\ref{key}) we write
\bea
& &\tQ_{-1,-1}+Q_{-1,0}\pa+\cdots+Q_{-1,n-2}\pa^{n-1}+Q_{-1,n-1}\pa^n\no\\
& &=-(\Q_{-1})'+\psi\Q_0+Q_{-1,n-1}(L_{(n,1)})_+
\eea
which implies
\be
\tQ_{-1,-1}=-\pa\Q_{-1}+\psi\Q_0+Q_{-1,n-1}(L_{(n,1)})_+
\lb{qtel}
\ee
Since a scalar is invariant under the adjoint operation; i.e. $f=f^*$, applying the
adjoint operation to the both sides of (\ref{qtel}) gives
\bea
\tQ_{-1,-1}&=&(\Q^*_{-1}\pa+(L_{(n,1)})^*_+Q_{-1,n-1}+\Q^*_0\psi)_0\no\\
&=&((L_{(n,1)})^*_+Q_{-1,n-1}+\Q^*_0\psi)_0
\lb{qdag}
\eea
Now combining (\ref{n2}), (\ref{q-1}) and (\ref{qdag}) yields
\bea
\de\psi=-N_{-1,0}&=&Q'_{-1,0}-\psi Q_{0,0}-u_0Q_{-1,n-1}+\psi Q_{-1,-1}\no\\
&=&-\tQ_{-1,-1}+\psi Q_{-1,-1}\no\\
&=&-((L_{(n,1)})^*_+Q_{-1,n-1}+\Q^*_0\psi)_0+\psi\int^x(\phi Q_{-1,n-1}+
\psi Q_{0,-1})\no\\
&=&-((L_{(n,1)})^*X^*\psi)_0-((L_{(n,1)})^*Q_{-1,n-1})_0+
\psi\int^x(\phi Q_{-1,n-1}+\psi Q_{0,-1})
\lb{deltapsi}
\eea
which is one of the desired Hamiltonian flow equations if we identify
\bea
X&=&\frac{\de H}{\de (L_{(n,1)})_+}=\pa^{-1}\frac{\de H}{\de u_0}+\cdots+
\pa^{-n}\frac{\de H}{\de u_{n-1}}\no\\
Q_{-1,n-1}&=&\frac{\de H}{\de \phi}
\lb{identify}\\
Q_{0,-1}&=&-\frac{\de H}{\de \psi}\no
\eea
The identification (\ref{identify}), which is a generalization of the one used for
GD hierarchy (\ref{eqln}) \cite{D1},  is motivated by the following relation
\bea
\de H&=& \int(\frac{\de H}{\de u_0}\de u_0+\cdots+
\frac{\de H}{\de u_{n-1}}\de u_{n-1}+\frac{\de H}{\de \phi}\de \phi+
\frac{\de H}{\de \psi}\de \psi)\no\\
&=&\Tr(Q\de{\cal{L}}_{(n,1)}) 
\lb{relationds}
\eea
where $\Tr(A)=\int\tr(A)$ and $\tr(A)$ denotes the ordinary trace of a square matrix. 
The relation (\ref{relationds}) is actually quite common in the matrix formulation
of an integrable hierarchy \cite{DS}. It is related to the fact that the operation 
$\Tr (A)$ provides a natural scalar product between the cotangent space
(to which $Q$ belongs) and the tangent space 
(to which $\de {\cal{L}}_{(n,1)}$ belongs) of the phase space manifold.
In the case of GD hierarchy the identification of this sort can be derived from
the associated linear system \cite{BLX}. However, we have not devised a similar
proof for the present case.
Here, the validity of (\ref{identify}) will be simply justified by the final result.

Now we treat $\tQ_{-1,-1}$ a little differently. An equivalent form of (\ref{qtel}) is
\be
\pa^{-1}\tQ_{-1,-1}=-\Q_{-1}+\pa^{-1}(\psi\Q_0+Q_{-1,n-1}(L_{(n,1)})_+)
\ee
which leads to
\be
\Q_{-1}=(\Q_{-1})_+=(\pa^{-1}\psi\Q_0+\pa^{-1}Q_{-1,n-1}L_{(n,1)})_+
\lb{qdown}
\ee
From (\ref{diff}) we have
\be
\Q^{-1}=[(Q_{0,-1}-X\phi)\pa^{-1}]_{\geq -n}
\lb{qup}
\ee
Now putting (\ref{qdown}), (\ref{qup}) into (\ref{lplus}) we obtain
\bea
\de (L_{(n,1)})_+&=&(L_{(n,1)}X)_+(L_{(n,1)})_+-
(L_{(n,1)})_+(XL_{(n,1)})_+-[L_{(n,1)}(Q_{0,-1}-X\phi)\pa^{-1}]_+\psi\no\\
& &-\phi(\pa^{-1}\psi\Q_0+\pa^{-1}Q_{-1,n-1}L_{(n,1)})_+
\eea
Using (\ref{q0}) and (\ref{identify}) we end up with
\bea
\de (L_{(n,1)})_+&=& (L_{(n,1)}X)_+(L_{(n,1)})_+-(L_{(n,1)})_+(XL_{(n,1)})_++
[((L_{(n,1)}X)_+\phi+L_{(n,1)}\frac{\de H}{\de \psi})\pa^{-1}\psi]_+\no\\
& &-[\phi\pa^{-1}\psi(XL_{(n,1)})_++\phi\pa^{-1}\frac{\de H}{\de \phi}L_{(n,1)}]_+\no\\
&=&[(L_{(n,1)}X)_+L_{(n,1)}-L_{(n,1)}(XL_{(n,1)})_+]_++
[L_{(n,1)}\frac{\de H}{\de \psi}\pa^{-1}\psi-
\phi\pa^{-1}\frac{\de H}{\de \phi}L_{(n,1)}]_+,
\lb{deltal}
\eea
an expected result.

Finally we turn to (\ref{n4}) and (\ref{dynaphi}) :
\be
\de \phi=N_{n-1,-1}=
-Q'_{n-1,-1}+\phi Q_{n-1,n-1}-\phi Q_{-1,-1}-\sum_{k=0}^{n-1}u_kQ_{k,-1}
\lb{vph}
\ee
Note first that (\ref{qi+1}) gives
\bea
Q_{1,-1}&=&Q'_{0,-1}-\phi Q_{0,n-1}\no\\
Q_{2,-1}&=&Q''_{0,-1}-\phi Q_{1,n-1}-(\phi Q_{0,n-1})'\no\\
&\cdots&\no\\
Q_{j,-1}&=&Q^{(j)}_{0,-1}-\phi Q_{j-1,n-1}-(\phi Q_{j-2,n-1})'\cdots-
(\phi Q_{0,n-1})^{(j-1)}
\lb{qa}\\
&\cdots&\no\\
Q_{n-1,-1}&=&Q^{(n-1)}_{0,-1}-\phi Q_{n-2,n-1}-(\phi Q_{n-3,n-1})'\cdots-
(\phi Q_{0,n-1})^{(n-2)}\no
\eea
In particular,
\be
Q'_{n-1,-1}=Q^{(n)}_{0,-1}-\phi Q_{n-1,n-1}-(\phi Q_{n-2,n-1})'-\cdots-
(\phi Q_{0,n-1})^{(n-1)}+\phi Q_{n-1,n-1}
\lb{qb}
\ee
The operator forms of (\ref{qa}) and (\ref{qb}) are, respectively, 
\bea
Q_{j,-1}&=&[\pa^j(Q_{0,-1}-X\phi)]_0\qquad (j=0,1,\cdots,n-1)
\lb{qc}\\
Q'_{n-1,-1}&=&[\pa^n(Q_{0,-1}-X\phi)]_0+\phi Q_{n-1,n-1}
\lb{qd}
\eea
Putting (\ref{qc}) and (\ref{qd}) into (\ref{vph}) we obtain the desired equation :
\bea
\de \phi&=&-[\pa^n(Q_{0,-1}-X\phi)]_0-
\sum_{k=0}^{n-1}u_j[\pa^j(Q_{0,-1}-X\phi)]_0-\phi Q_{-1,-1}\no\\
&=&[(L_{(n,1)})_+(X\phi+\frac{\de H}{\de \psi})]_0-
\phi\int^x(\phi\frac{\de H}{\de \phi}-\psi\frac{\de H}{\de \psi})\no\\
&=&(L_{(n,1)}X\phi)_0+(L_{(n,1)}\frac{\de H}{\de \psi})_0-
\phi\int^x(\phi\frac{\de H}{\de \phi}-\psi\frac{\de H}{\de \psi})
\lb{deltaphi}
\eea
If we remember identification, $\de f\equiv \{f, H\}$ mentioned in the paragraph
following (\ref{dynau}), where $f$ is either of the dynamical variables 
$u_i$, $\phi$ and $\psi$, then
eqs. (\ref{deltapsi}), (\ref{deltal}) and (\ref{deltaphi}) together define the second 
Hamiltonian structure of the one-constraint KP hierarchy.
In conclusion, we have shown that the bracket of Oevel and Strampp \cite{OS}
comes out naturally from the matrix formulation 
(which, of course, is basically the AKNS scheme).

\section{Generalizations}

Having derived the Hamiltonian structure of the one-constraint KP hierarchy from a matrix 
formulation we now come to the multi-constraint case.
The Lax operator for this case is defined by
\be
L_{(n,m)}=\pa^n+u_{n-1}\pa^{n-1}+\cdots+u_0+\sum_{j=1}^{m}\phi_j\pa^{-1}\psi_j
\lb{laxlnm}
\ee
which has matrix representation of the form
\bea
{\cal{L}}_{(n,m)}&=&I\pa+U\\
U&=&
\left(
\ba{cccccccc}
0 &\cdots & 0 & -\psi_m& \cdots &\cdots  &\cdots  & 0\\
 & \ddots & \ddots&\vdots & & & & \\
 & & 0 &-\psi_1 & & & \\
& &  & 0 & -1 &  &  &\\
& &  &  & \ddots & -1 &  & \\
 & & &  & & \ddots & \ddots & \\
& & & & & & 0 & -1\\
\phi_m &\cdots &\phi_1 & u_0 & u_1 & \cdots & \cdots &u_{n-1}
\ea
\right)
\eea
where $U$ is a $(n+m)\times(n+m)$ matrix and the numeration 
of rows and columns will be $i, j=-m,-m+1,\cdots,n-1$.
Our matrix is connected by the Miura transformation, within the generalized
Wilson-Drinfeld-Sokolov method, with the matrix used in ref.\cite{AFGZ}.
We also introduce another $(n+m)\times(n+m)$ matrix $Q$
such that $[Q,{\cal{L}}_{(n,m)}]$ consistent with the form of $\de{\cal{L}}_{(n,m)}$.
We can compute the matrix elements $N_{ij}\equiv {[Q,{\cal{L}}_{(n,m)}]}_{ij}$
and make it of the form $\de{\cal{L}}_{(n,m)}$. 
The result is the following :\\
(1) $i< 0$ or $j<0$
\bea
N_{-\al,-\bt}&=&-Q'_{-\al,-\bt}+\phi_\bt Q_{-\al,n-1}+\psi_\al Q_{0,-\bt}\\
N_{-\al,0}&=&-Q'_{-\al,0}+u_0Q_{-\al,n-1}+\psi_\al Q_{0,0}-
\sum_{k=1}^m\psi_kQ_{-\al,-k}\\
N_{-\al,i}&=&-Q'_{-\al,i}-Q_{-\al,i-1}+u_iQ_{-\al,n-1}+\psi_{\al}Q_{0,i}
\qquad (i=1,2,\cdots,n-1)\\
N_{i,-\al}&=&-Q'_{i,-\al}+\phi_{\al}Q_{i.n-1}+Q_{i+1,-\al}
\qquad (i=0,1,\cdots,n-2)\\
N_{n-1,-\al}&=&-Q'_{n-1,-\al}+\phi_{\al}Q_{n-1,n-1}-\sum_{k=1}^m\phi_k
Q_{-k,-\al}-\sum_{k=0}^{n-1}u_kQ_{k,-\al}
\eea
where $\al, \bt=1,2,\cdots,m$.\\
(2) $i\geq 0$ and $j\geq 0$
\bea
N_{i,0}&=&-Q'_{i,0}+u_0Q_{i,n-1}+Q_{i+1,0}-\sum_{k=1}^m\psi_kQ_{i,-k}
\qquad (i=0,1,\cdots,n-2)\\
N_{n-1,0}&=&-Q'_{n-1,0}+u_0Q_{n-1,n-1}-\sum_{k=1}^m\psi_kQ_{n-1,-k}
-\sum_{k=1}^m\phi_kQ_{-k,0}-\sum_{k=0}^{n-1}u_kQ_{k,0}\\
N_{i,j}&=&-Q'_{i,j}-Q_{i,j-1}+u_jQ_{i,n-1}+Q_{i+1,j}\qquad
{{i=0,1,\cdots,n-2} \choose {j=1,2,\cdots,n-1}}\\
N_{n-1,j}&=&-Q'_{n-1,j}-Q_{n-1,j-1}+u_jQ_{n-1,n-1}-\no\\
& &\sum_{k=1}^m\phi_kQ_{-k,j}-\sum_{k=0}^{n-1}Q_{k,j}\qquad
(j=1,2,\cdots,n-1)\\
\eea
with 
\bea
N_{-j,0}&=&-\de \psi_j \\
N_{n-1,-j}&=&\de \phi_j\qquad j=1,2,\cdots,m\\
N_{n-1,j}&=&\de u_j\qquad j=0,1,\cdots,n-1\\
N_{i,j}&=&0\qquad \mbox{otherwise}
\eea
The definition of $\Q_i$, $\Q^i$ and $\hat{N}_i$  
are still given by (\ref{defq}) and (\ref{defn}).
Following the steps presented in the previous section and imposing
the following identifications
\bea
X&\equiv&\Q^{n-1}=\frac{\de H}{\de (L_{(n,m)})_+}\no\\
Q_{-k,n-1}&=&\frac{\de H}{\de \phi_k}\\
Q_{0,-k}&=&-\frac{\de H}{\de \psi_k}\qquad (k=1,2,\cdots,m)\no
\eea
we obtain after performing a similar derivation the Hamiltonian structure associated with the 
Lax operator (\ref{laxlnm}) :
\bea
\de (L_{(n,m)})_+&=&
[(L_{(n,m)}X)_+L_{(n,m)}-L_{(n,m)}(XL_{(n,m)})_+]_+\no\\
& &+\sum_{k=1}^m[L_{(n,m)}\frac{\de H}{\de \psi_k}\pa^{-1}\psi_k-
\phi_k\pa^{-1}\frac{\de H}{\de \phi_k}L_{(n,m)}]_+
\lb{deltalnm}\\
\de\phi_j&=&
(L_{(n,m)}X\phi_j)_0+(L_{(n,m)}\frac{\de H}{\de \psi_j})_0-\sum_{k=1}^m
\phi_k\int^x(\phi_j\frac{\de H}{\de \phi_k}-\psi_k\frac{\de H}{\de \psi_j})\\
\de \psi_j&=&
-((L_{(n,m)})^*X^*\psi_j)_0-((L_{(n,m)})^*\frac{\de H}{\de \phi_j})_0+\sum_{k=1}^m
\psi_k\int^x(\phi_k \frac{\de H}{\de \phi_j}-\psi_j \frac{\de H}{\de \psi_k})
\lb{deltapsinm}
\eea
We would like to remark that the Hamiltonian flow equations 
(\ref{deltalnm})-(\ref{deltapsinm}) are, in fact, coming from the second 
GD brackets defined by the Lax operator (\ref{laxlnm}).
To see this, let us denote $\frac{\de H}{\de L_{(n,m)}}=
\frac{\de H}{\de (L_{(n,m)})_+}+A$ where
 $A$ is a differential operator. From the identity
\be
\tr(\frac{\de H}{\de L_{(n,m)}}\de L_{(n,m)})=
\tr(\frac{\de H}{\de (L_{(n,m)})_+}\de (L_{(n,m)})_+)+\sum_{j=1}^m
\int(\frac{\de H}{\de \phi_j}\de \phi_j+\frac{\de H}{\de \psi_j}\de \psi_j)
\ee
we find that the differential operator $A$ satisfies
\be
(A\phi_j)_0=\frac{\de H}{\de \psi_j}\qquad (A^*\psi_j)_0=\frac{\de H}{\de \phi_j}
\lb{cond}
\ee
Now putting $L_{(n,m)}$ and $\frac{\de H}{\de L_{(n,m)}}=
\frac{\de H}{\de (L_{(n,m)})_+}+A$
into the second GD bracket (\ref{gd2}) and using the relation (\ref{cond}), we obtain
the Poisson brackets (\ref{deltalnm})-(\ref{deltapsinm}).

Note that when we set the next leading coefficient $u_{n-1}$ to vanish we have to impose
the constraint $\res[\frac{\de H}{\de L_{(n,m)}}, L_{(n,m)}]=0$. Then
the resulting bracket becomes the modified GD bracket (\ref{mgd2}) defined by
$L_{(n,m)}$ with $u_{n-1}=0$. From this modified bracket,
it can be shown \cite{Liu} that the Hamiltonian structures of the coupled AKNS
hierarchy ($n=1,m=2$) and the coupled Yajima-Oikawa 
hierarchy \cite{YO} ($n=2, m=2$) come out as the special cases.

In the remaining part of this section, we would like to consider another
reduction of the KP hierarchy which has Lax operator of the form
\be
K=\pa^n+u_{n-1}\pa^{n-1}+\cdots+u_0+a\frac{1}{\pa-s}
\lb{laxk}
\ee
Such system was introduced in the study of matrix model \cite{BX} and can be
written in the form (\ref{laxln1}) if we equate with each other the integral parts
of the two Lax operators :
\be
\phi\pa^{-1}\psi=a\frac{1}{\pa-s}
\ee
Then
\be
a=\phi\psi\qquad s=-(\ln\psi)'
\lb{gt1}
\ee
or, equivalently
\be
\phi=ae^{\int^xs}\qquad \psi=e^{-\int^xs}
\ee
We can also formulate the Hamiltonian structure associated with $K$
in matrix form by introducing
\be
{\cal{K}}=
\left(
\ba{cccccc}
\pa-s & -1 & 0 & 0 & \cdots & 0 \\
\cdots & \pa & -1 & 0 & \cdots  & \cdots \\
\cdots &\cdots   & \cdots & -1 &\cdots & 0\\
 \cdots & \cdots  & \cdots &\cdots  & \cdots & 0\\
\cdots & \cdots &\cdots  & \cdots & \pa & -1\\
a & u_0 & u_1 & \cdots & \cdots &\pa+u_{n-1}
\ea
\right)
\ee
and choosing a $(n+1)\times(n+1)$ matrix $R$
such that $[R,{\cal{K}}]$ consistent with $\de {\cal{K}}$.

After equating the matrix elements of $[R,{\cal{K}}]$ with the matrix elements of 
$\de{\cal{K}}$:
\bea
\de {\cal{K}}_{-1,-1}&=&-\de s
\lb{dynas}\\
\de {\cal{K}}_{n-1,-1}&=&\de a\\
\de {\cal{K}}_{n-1,j}&=&\de u_j\qquad (j=0,1,\cdots,n-1)
\lb{dynauj}\\
\de {\cal{K}}_{i,j}&=&0\qquad \mbox{otherwise}
\lb{const}
\eea
we can solve $R_{ij}$ in terms of $R_{-1,-1}$ and $R_{i,n-1}$ from
the constraint equation (\ref{const}).
Thus the remaining task is to identify $R_{-1,-1}$ and $R_{i,n-1}$.
However, the following naive identifications
\bea
R_{-1,-1}&=& -\frac{\de H}{\de s}\\
R_{-1,n-1}&=& \frac{\de H}{\de a}
\lb{ida}\\
R_{i,n-1}&=&\frac{\de H}{\de u_i}\qquad (i=0,1,\cdots,n-1)
\lb{idu}
\eea
do not give us the correct Hamiltonian structure. To get correct identifications we
note that Eq.(\ref{gt1}) can be expressed  as a matrix equation:
\be
{\cal{K}}=\Psi^{-1}{\cal{L}}_{(n,1)}\Psi
\lb{matrixgt}
\ee
where the matrix $\Psi$ is defined by
\be
\left(
\ba{cccc}
\psi & \cdots& \cdots& 0\\
& 1 & & \\
& &\ddots & \\
0&\cdots &\cdots & 1
\ea
\right)
\ee

From (\ref{matrixgt}) and the Hamiltonian flow equation 
$\de{\cal{L}}_{(n,1)}=[Q,{\cal{L}}_{(n,1)}]$ 
it is easy to show that $\de{\cal{K}}=[-\Psi^{-1}\de\Psi+\Psi^{-1}Q\Psi, {\cal{K}}]$. 
As a consequence, $R$ is related to $Q$ determined in the previous section by
\be
R=-\Psi^{-1}\de\Psi+\Psi^{-1}Q\Psi
\ee
From this relation we find that the correct identification of  $R_{-1,-1}$ should be
\be
R_{-1,-1}= -\frac{\de H}{\de s}+\pa^{-1}_x\de s
\lb{ids}
\ee

Substituting (\ref{ida}), (\ref{idu}) and (\ref{ids}) into 
the ``dynamical" equations, (\ref{dynas})-(\ref{dynauj}), 
involving $\de a$, $\de s$, and $\de L_+$ we obtain the Hamiltonian structure
\bea
\de K_+&=& (K_+X)_+K_+-K_+(XK_+)_++
[K_+(Xa+a\frac{\de H}{\de a}+(\frac{\de H}{\de s})')(\pa-s)^{-1}]_+\no\\
& & -a[(\pa-s)^{-1}(X+\frac{\de H}{\de a})K_+]_+
\lb{deltak}\\
\de a&=&[e^{-\int^x s}K_+(Xa+a\frac{\de H}{\de a}+
(\frac{\de H}{\de s})')e^{\int^x s}]_0-
a[e^{\int^x s}K^*_+(X^*+\frac{\de H}{\de a})e^{-\int^x s}]_0 \\
\de s&=&(\frac{\de H}{\de s})'+
[e^{\int^x s}K^*_+(X^*+\frac{\de H}{\de a})e^{-\int^x s}]'_0
\lb{deltas}
\eea
where $X\equiv \frac{\de H}{\de K_+}$.
One can show by explicit calculations that (\ref{deltak})-(\ref{deltas}) is indeed 
the same as the second GD bracket defined by $K$\cite{BLX}.

Rigorously to say, what we obtain here is not really a matrix formulation of the second
Hamiltonian structure associated with $K$. This reason is that the identification
of $R_{-1,-1}$, (\ref{ids}), contains the term $\pa_x^{-1}\de s$ which can be 
determined only after $\de s$ is computed. In other words, $R$ is not really the
``dual space" of the matrix ${\cal{K}}$ (which contains all dynamical variables).
The question whether or not a matrix formulation for the second GD bracket defined by 
$K$ remains open.

\section{Examples}

In this section, we consider two explicit examples. The first is the cKP hierarchy 
associated with the Lax operator
\be
L_{(2,1)}=\pa^2+u+\phi\pa^{-1}\psi.
\ee
It is easy to transform the corresponding eigenvalue equation $L_{(2,1)}\varphi=0$ into
a matrix form by taking
\be
{\cal{L}}_{(2,1)}=
\left(
\ba{ccc}
\pa & -\psi & 0 \\
0 & \pa & -1 \\
\phi & u & \pa
\ea
\right)
\qquad
\Phi=
\left(
\ba{c}
\varphi_1\\
\varphi_2\\
\varphi_3
\ea
\right)
\ee
Then ${\cal{L}}_{(2,1)}\Phi$=0 implies
\bea
& &\vphi'_1-\psi\vphi_2=0
\lb{eq1}\\
& &\vphi'_2-\vphi_3=0
\lb{eq2}\\
& &\phi\vphi_1+u\vphi_2+\vphi'_3=0
\lb{eq3}
\eea
Substitutions of (\ref{eq1}) and (\ref{eq2}) into (\ref{eq3}) gives
\be
(\pa^2+u+\phi\pa^{-1}\psi)\vphi_2=0
\ee
which is precisely $L_{(2,1)}\vphi=0$ once $\vphi=\vphi_2$ is imposed.

To consider the Hamiltonian structure associated with ${\cal{L}}_{(2,1)}$, let us
consider the associated Lax equation
\be
\pa_t{\cal{L}}_{(2,1)}=[{\cal{M}},{\cal{L}}_{(2,1)}]
\ee
It is necessary to solve the form of ${\cal{M}}$ to make the above
equation consistent.

Writing
\be
{\cal{M}}=
\left(
\ba{ccc}
a & b & c\\
d & e & f\\
g & h & -a-e
\ea
\right)
\ee
we have the following explicit expression for the commutator 
$[{\cal{M}}, {\cal{L}}_{(2,1)}]$:
\be
{\cal{M}}{\cal{L}}_{(2,1)}-{\cal{L}}_{(2,1)}{\cal{M}}=
-{\cal{M}}'+
\left(
\ba{ccc}
c\phi+\psi d & -a\psi+cu+\psi e & -b+\psi f\\
f\phi+g & -d\psi+fu+h & -a-2e\\
-(2a+e)\phi-ud, & -g\psi-(a+2e)u-\phi b, & -h-c\phi-fu
\ea
\right)
\lb{comt}
\ee
Setting the matrix (\ref{comt}) to equal to
\be
\de{\cal{L}}_{(2,1)}=
\left(
\ba{ccc}
0 & -\de\psi & 0\\
0 & 0 & 0\\
\de\phi & \de u & 0
\ea
\right)
\ee
we obtain five constraint equations
\bea
& &-a'+c\phi+\psi d=0\no\\
& &-c'-b+\psi f=0\no\\
& &-d'+f\phi+g=0\no\\
& &-e'-d\psi+fu+h=0\no\\
& &-f'-2e-a=0
\eea
It is easy to see that $a,b,e,g,h$ can be solved in terms of $c,d,f$ and $u,\phi,\psi$
\bea
a&=&\pa^{-1}(c\phi+\psi d)\no\\
b&=&-c'+\psi f\no\\
g&=&d'-f\phi\\
e&=&-\frac{1}{2}(f'+\pa^{-1}(c\phi+\psi d))\no\\
h&=&-\frac{1}{2}(f''+c\phi-d\psi)-fu
\eea
On the other hand, we have
\bea
\de u&=&-h'-g\psi-(a+2e)u-\phi b\\
\de\phi&=&-g'-(2a+e)\phi-ud\\
\de\psi&=&b'+(a-e)\psi-cu
\eea
Hence
\bea
\de u&=&
(\frac{1}{2}\pa^3+\pa u+u\pa)f+\frac{1}{2}(\pa\phi+\phi\pa)c-
\frac{1}{2}(\pa\psi+\psi\pa)d\\
\de\phi&=&
\frac{1}{2}(\pa\phi+\phi\pa)f-\frac{3}{2}(\phi\pa^{-1}\phi)c-
(\pa^2+u+\frac{3}{2}\phi\pa^{-1}\psi)d\\
\de\psi&=&
\frac{1}{2}(\pa\psi+\psi\pa)f-(\pa^2+u-\frac{3}{2}\psi\pa^{-1}\phi)c+
\frac{3}{2}(\psi\pa^{-1}\psi)d
\eea
If we identify
\be
f=\frac{\de H}{\de u}\qquad c=\frac{\de H}{\de \phi}
\qquad d=-\frac{\de H}{\de \psi}
\ee
and regard $\de u$, $\de \phi$ and $\de \psi$ as Hamiltonian flows, then we can read
off the Poisson brackets
\bea
\{ u(x), u(y)\}&=& [\frac{1}{2}\pax^3+\pax u(x)+u(x)\pax]\de(x-y)\\
\{ u(x), \phi(y)\}&=& [\phi(x)\pax+\frac{1}{2}\phi'(x)]\de(x-y)\\
\{ u(x), \psi(y)\}&=& [\psi(x)\pax+\frac{1}{2}\psi'(x)]\de(x-y)\\
\{ \phi(x), \phi(y)\}&=& -\frac{3}{2}\phi(x)\ep(x-y)\phi(y)\\
\{ \phi(x), \psi(y)\}&=& [\pax^2+u(x)]\de(x-y)+\frac{3}{2}\phi(x)\ep(x-y)\psi(y)\\
\{ \psi(x), \psi(y)\}&=& -\frac{3}{2}\psi(x)\ep(x-y)\psi(y)
\eea
where $\ep(x-y)\equiv\pax^{-1}\de (x-y)$ is the antisymmetric step function.
These are correct brackets\cite{OS}, which can be computed from the modified 
GD bracket (\ref{mgd2}) associated with $L=\pa^2+u+\phi\pa^{-1}\psi$.

The second example is to consider the Lax operator of the form
\be
K=\pa^2+u+a\frac{1}{\pa-s}
\ee
Even though we would not get a genuine matrix formulation of the second 
GD bracket associated with $K$ as discussed in the previous section. However,
we like to check explicitly for this simple case that our identifications
(\ref{ida}), (\ref{idu}) and (\ref{ids}) indeed give us a correct Hamiltonian structure.

It is also easy to derive a matrix representation for $K$
\be
{\cal{K}}=
\left(
\ba{ccc}
\pa-s & -1 & 0\\
0 & \pa & -1 \\
a & u &\pa
\ea
\right)\qquad
\Phi=
\left(
\ba{c}
\phi_1\\
\phi_2\\
\phi_3
\ea
\right)
\ee
Then from ${\cal{K}}\Phi=0$ we have
\bea
& &(\pa-s)\phi_1-\phi_2=0\\
& &\phi_2'-\phi_3=0\\
& &a\phi_1+u\phi_2+\phi'_3=0
\eea
which imply
\be
[\pa^2+u\phi_2+a(\pa-s)^{-1}]\phi_2=0
\ee
as desired. The associated Lax equation is given by
\be
\pa_t\cal{K}=[\cal{N},\cal{K}]
\ee
Writing
\be
\cal{N}=
\left(
\ba{ccc}
 n_1 & n _2 & n_3 \\
 n_4 & n_5 & n_6\\
n_7 & n_8 &n_9
\ea
\right)
\ee
then
\be
[\cal{N},\cal{K}]=
-\cal{N}'+
\left(
\ba{ccc}
 an_3+n_4&-n_1+un_3+sn_2+n_5&-n_2+sn_3+n_6 \\
-sn_4+an_6+n_7&-n_4+un_6+n_8 & -n_5+n_9\\
-sn_7+an_9-an_1-un_4 &-n_7+un_9-an_2-un_5 &-n_8-an_3-un_6
\ea
\right)
\ee
which must be equal to
\be
\de\cal{K}=
\left(
\ba{ccc}
 -\de s & 0 & 0 \\
 0 & 0 & 0\\
\de a & \de u & 0
\ea
\right)
\ee
There are six constraint equations
\bea
& &-n'_2-n_1+un_3+sn_2+n_5=0
\lb{kn2}\\
& &-n'_3-n_2+sn_3+n_6=0\\
& &-n'_4-sn_4+an_6+n_7=0\\
& &-n'_5-n_4+un_6+n_8=0\\
& &-n'_6-n_5+n_9=0\\
& &-n'_9-n_8-an_3-un_6=0
\lb{kn9}
\eea
and three dynamical equations
\bea
\de s&=& n'_1-an_3-n_4
\lb{ks}\\
\de a&=& -n'_7-sn_7+an_9-an_1-un_4\\
\de u&=&-n'_8-n_7+un_9-an_2-un_5
\lb{ku}
\eea
We can solve $n_2, n_4, n_5, n_7, n_8$, and $n_9$ from 
(\ref{kn2})-(\ref{kn9}) in terms of $n_1, n_3$, and $n_6$ as follows
\bea
n_2&=&-(\pa-s)n_3+n_6
\lb{solvn2}\\
n_4&=&(2\pa(\pa-s)^2+2\pa u-a)n_3-2\pa n_1-\pa(3\pa-2s)n_6\\
n_5&=&-((\pa-s)^2+u)n_3+(\pa-s)n_6+n_1\\
n_7&=&(2(\pa+s)\pa(\pa-s)^2+2(\pa+s)\pa u-(\pa+s)a))n_3\no\\
& &-2(\pa+s)\pa n_1-((\pa+s)\pa(3\pa-2s)+a)n_6\\
n_8&=&(\pa(\pa-s)^2+\pa u-a)n_3-(2\pa^2-\pa s+u)n_6-\pa n_1\\
n_9&=&-((\pa-s)^2+u)n_3+(2\pa-s)n_6+n_1
\lb{solvn9}
\eea
 Now substituting (\ref{solvn2})-(\ref{solvn9}) into the dynamical equations 
(\ref{ks})-(\ref{ku}) and using the identifications
\be
n_1=-\frac{\de H}{\de s}+\pax^{-1}\de s,\qquad n_3=\frac{\de H}{\de a}\qquad
n_6=\frac{\de H}{\de u}
\ee
we obtain the expected Poisson brackets\cite{BLX}
\bea
\{u(x), u(y)\}&=&[\frac{1}{2}\pax^3+2u(x)\pax+u'(x)]\de(x-y)\\
\{u(x), a(y)\}&=&[3a(x)\pax+2a'(x)]\de(x-y)\\
\{u(x), s(y)\}&=&[\frac{3}{2}\pax^2+s(x)\pax]\de(x-y)\\
\{a(x), a(y)\}&=&[(4a(x)s(x)+2a'(x))\pax+a''(x)+2(a(x)s(x))']\de(x-y)\\
\{a(x), s(y)\}&=&[(\pax+s(x))^2\pax+u(x)\pax]\de(x-y)\\
\{s(x), s(y)\}&=&\frac{3}{2}\pax\de(x-y)
\eea

\section{Concluding Remarks}

We have studied the Hamiltonian structure of the cKP hierarchy
in the matrix formulation. We showed that the procedure for the GD
hierarchy can be gone through without difficulty for the cKP hierarchy.
We have not only reproduced Oevel and Strampp's result for the one-constraint
case but also generalized it to the multi-constraint case. We further showed
that this Hamiltonian structure is nothing but the GD bracket.
Hence our result again confirms the fact that the GD bracket can be properly
restricted to the Lax operator of the form (\ref{laxlnm}).
We have also discussed the matrix formulation of the Hamiltonian
structure of the one-constraint KP hierarchy in the form given by
(\ref{laxk}). However, we did not obtain a genuine matrix formulation
in the usual sense. Nevertheless, we found that one can still compute
the Hamiltonian structure from our matrix formulation once a proper modification
is made.

Finally, we want to remark that the constrained modified KP hierarchy\cite{OS}
can be obtained from the Lax operator (\ref{laxln1}) via the gauge transformation
\bea
K&=& \phi^{-1}L_{(n,1)}\phi\no\\
&=&\pa^n+v_{n-1}+\cdots+v_0+\pa^{-1}v_{-1}
\eea
which satisfies the hierarchy equation
\be
\frac{\pa K}{\pa t_k}=[K^n_{\geq 1},K]
\lb{eqcmkp}
\ee
The bi-Hamiltonian structure associated with (\ref{eqcmkp}) has been obtained \cite{OS}
from the lifted bracket of the cKP hierarchy by gauge transformation.
It would be interesting to know whether or not a matrix formulation of this
bi-Hamiltonian structure exists. Following the spirit of the present work
we  have worked out this structure in matrix formulation for $n=2$ and $3$,
but a general construction is yet to be given.
We hope to report the results in this direction in the near future.

{\bf Acknowledgments\/}
This work is supported by the National Science Council of Taiwan 
under grant No. NSC-87-2112-M-018-008 (W.J.H) and 
NSC-87-2811-M-007-0025 (J.C.S and M.H.T).
One of us (M.H.T) also wishes to thank Center for Theoretical Sciences of National
Science Council of Taiwan for partial support.\\

\end{document}